\documentclass{cernrep}
\usepackage{graphicx,here}

\def\be{\begin{equation}}
\def\ee{\end{equation}}

\begin{document}

\title{Energetic Cosmic Rays observed by the resonant gravitational 
wave detector NAUTILUS}

\author{
P. Astone$^1$, M. Bassan$^2$, P. Bonifazi$^3$, P. Carelli$^4$,\\
E. Coccia$^2$, S.D'Antonio$^5$,
V. Fafone$^6$, G.Federici$^1$,\\ A. Marini$^6$, G. Mazzitelli$^6$,
Y. Minenkov$^2$, I. Modena$^2$,\\
G. Modestino$^6$,
A. Moleti$^2$, G. V. Pallottino$^5$, V. Pampaloni$^6$ \\ G. Pizzella$^7$,
L.Quintieri$^6$, F. Ronga$^6$, R. Terenzi$^3$, 
M. Visco$^3$, L. Votano$^6$ \\
$~$\\
$~$
}

\vskip 0.1 in

\institute{
{\it ${}^{1)}$ Istituto Nazionale di Fisica Nucleare INFN, Rome}\\
{\it ${}^{2)}$ University of Rome "Tor Vergata" and INFN, Rome}\\
{\it ${}^{3)}$ IFSI-CNR and INFN, Frascati}\\
{\it ${}^{4)}$ University of L'Aquila and INFN, Rome}\\
{\it ${}^{5)}$ University of Rome "La Sapienza" and INFN, Rome}\\
{\it ${}^{6)}$ Istituto Nazionale di Fisica Nucleare INFN, Frascati}\\
{\it ${}^{7)}$ University of Rome "Tor Vergata" and INFN, Frascati}
}

\maketitle
 
\begin{abstract}
Cosmic ray showers interacting with the resonant mass gravitational wave antenna
NAUTILUS have been detected. The experimental results show large signals at a
rate much greater than expected.  The largest signal corresponds to an energy
release in NAUTILUS of 87 TeV. We remark that a resonant mass gravitational wave
detector used as particle detector has characteristics different from the usual
particle detectors, and it could detect new features of cosmic rays. Among
several possibilities, one can invoke unexpected behaviour of superconducting
Aluminium as particle detector, producing enhanced signals, the excitation of
non-elastic modes with large energy release or anomalies in cosmic rays (for
instance, the showers might include exotic particles as nuclearites or Q-balls).
Suggestions for explaining these observations are solicited.

\end{abstract}

%{\bf~~~~June 2000~~~~For circulation only within the ROG group}

%\pagestyle{plain}
%\setcounter{page}2
%\baselineskip=17pt
\section{Introduction}
\begin{figure}
 \vspace{9.0cm}
\includegraphics{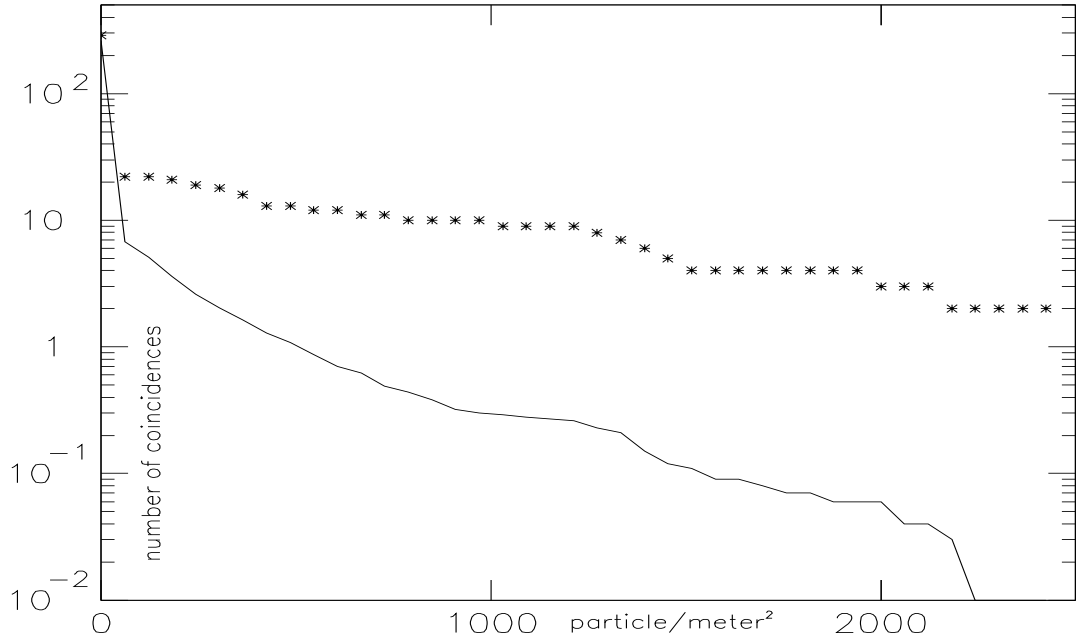}
 \caption{
Coincidences between the g.w. detector NAUTILUS and the c.r. detector. The
asterisks show the integral number of observed coincidences versus the particle
density observed by the c.r. counters located under the NAUTILUS cryostat. The
continuous line shows the estimated number of accidental coincidences.
        \label{versus} }
\end{figure}

The gravitational wave (g.w.) detector NAUTILUS has recently proven to be
capable of recording signals due to the passage of cosmic rays
\cite{prl}. In the ongoing
analysis of the data obtained with NAUTILUS in coincidence with cosmic ray
(c.r.) detectors we found new interesting results, which we are going to report
here. The work initially done by Beron and Hofstander
\cite{beron,beron1}, Strini and
Tagliaferri \cite{strini} and refined calculations by 
several authors \cite{allega,bernard,deru,amaldi,bari} estimated the
possible acoustic effects due to the passage of particles in a metallic bar. It
was predicted that for the vibrational energy in the longitudinal fundamental
mode of a metallic bar with length $L$ the following formula holds:
\be
E\simeq\frac{4}{9\pi}\frac{\gamma^2}{\rho L v^2}(\frac{dW}{dx})^2
\label{desudx}
\ee
where $E$ is the energy of the excited vibration mode, $\frac{dW}{dx}$
 is the energy loss
of the particle in the bar, $\rho$ is the density, $v$ the sound velocity in the
material and $\gamma$ is the Gr$\ddot{u}$neisen
 coefficient (depending on the ratio of the
material thermal expansion coefficient to the specific heat) which is commonly
considered constant with temperature. The adopted mechanism assumes that the
mechanical vibrations originate from the local thermal expansion caused by the
warming up due to the energy lost by the particles crossing the material. The
above formula has been recently verified by an experiment at room temperature
\cite{van},
using a small Aluminium cylinder and an electron beam. We notice that the g.w.
bar used as particle detector has characteristics very different from the usual
particle detectors, because the usual detectors are sensitive only to ionization
losses.
The resonant-mass g.w. detector NAUTILUS \cite{nau},
 operating at the INFN Frascati
Laboratory, consists of an aluminium 2300-kg bar cooled at 140 mK, below the
superconducting transition temperature \cite{coccia} of 0.92 K.
 Applying eq.\ref{desudx} to the case of NAUTILUS we find
\be
E=7.64~10^{-9}~W^2~f
\label{evw}
\ee
where $E$ is expressed in kelvin units , $W$ in GeV units is
 the energy delivered by
the particle to the bar and $f$  is a geometrical factor of the order of unity.
The bar and a resonant transducer, providing the read-out, form a coupled
oscillator system with two resonant modes, whose frequencies are
 $906.40~ Hz$ and $921.95 ~Hz$.
 The transducer converts the mechanical vibrations into an electrical
signal and is followed by a dcSQUID electronic amplifier. The NAUTILUS data,
recorded with a sampling time of $4.54~ms$, are processed by a
filter \cite{veloce} optimized to detect impulse 
signals applied to the bar, such as those due to a short burst
of g.w. In the present data analysis we consider antenna events defined as
follows. We apply to the filtered data a threshold corresponding to signal to
noise ratio $SNR = 19.5$, and for each threshold crossing we take the maximum
value above threshold and its time of occurrence. These two quantities define
the event of the g.w. detector. We wish to stress that here we consider only
events with energy greater than about twenty times the noise. The events
produced by NAUTILUS are posted on the WEB within the IGEC collaboration among
the groups that operate resonant g.w. detectors \cite{prodi}.
NAUTILUS is equipped with a c.r. detector system consisting of seven layers of
streamer tubes for a total of 116 counters \cite{rcd}.
 Three superimposed layers, each one with area of $36~m^2$, 
are located over the cryostat. Four superimposed layers
are under the cryostat, each one with area of $16.5~m^2$.
 Each counter measures the
charge, which is proportional to the number of particles. The detector is able
to measure particle density up to $5000 \frac{particles}{m^2}$
 without large saturation
effects and it gives a rate of showers in good agreement with the expected
number \cite{rcd,cocco},
 as verified here using the up particle density, which is not
affected by the interaction in the NAUTILUS detector.
We have searched for coincidences between the NAUTILUS events and the signals
from the c.r. NAUTILUS detectors in the period from 11 September 1998 until the
end of the year 1998, for a total observation time of $83.4~days$ where we have
26466 NAUTILUS events and 94775 c.r. events. We have determined a) the number of
coincidences, using a time window \cite{prl} of $0.5~s$,
 as a function of the particle
density of the c.r. events, b) the corresponding background of accidental
coincidences estimated by performing one hundred time shifts of the NAUTILUS
event times, in steps of 2 seconds. The result of the analysis, i.e. the number
nc of observed coincidences and the estimated number n of accidental
coincidences  versus the particle density is given in fig.\ref{versus}.

Clear coincidence excess above background is found, when the showers have
particle density large enough to give a signal in the bar. The eighteen
coincidences obtained for the down particle density  greater than 300
$\frac{particles}{m^2}$ with with expected number of accidentals $n=2.1$
are shown in table 1.

\begin{table}
\centering
\caption{
 List of eighteen coincidences between NAUTILUS and the c.r. detector
}
\vskip 0.1 in
\begin{tabular}{|c|c|c|c|c|c|c|c|}
\hline
day&hour&min&s&energy of the&noise of the&up particle&down particle\\
&&&&event&g.w.detector&density&density\\
&&&&$[K]$&$T_{eff}$ in $mK$&$[m^{-2}]$&$[m^{-2}]$\\
\hline

262&	23&	11&	29.581&	2.28	&0.003&	37&	312\\
277&	22&	26&	35.771&	0.04&	0.002&	118&	405\\
285&	17&	23&	14.9779&0.06&	0.002&	1238&	2494\\
286&	0&	35&	23.9222	&57.89&	0.004&	2442&	3556\\
295&	21&	0&	34.3376&0.07&	0.003&	235&	536\\
297&	21&	38&	49.9765	&0.37&	0.011&	547&	1374\\
303&	10&	38&	36.5147&0.42&	0.016&	227&	360\\
306&	8&	19&	59.5765	&0.12&	0.006&	629&	1409\\
311&	15&	24&	27.1148&0.12&	0.003&	751&	390\\
311&	15&	26&	21.0289	&0.14&	0.004&	148&	623\\
311&	23&	22&	8.4868&	0.45&	0.021&	223&	407\\
324&	14&	14&	47.3926&1.14&	0.044&	258&	785\\
350&	20&	56&	18.6130	&0.22&	0.004&	392&	1323\\
354&	23&	54&	19.2230&0.37&	0.004&	1064&	1972\\
356&	3&	17&	35.7440	&0.09&	0.004&	434&	2169\\
358&	0&	19&	21.9564&0.04&	0.002&	286&	1234\\
361&	12&	49&	13.9211	&0.09&	0.003&	258&	983\\
365&	12&	35&	40.6593&0.32&	0.007&	324&	1490\\
\hline
\end{tabular}
\label{eventi}
\end{table}

\begin{figure}
 \vspace{9.0cm}
\includegraphics{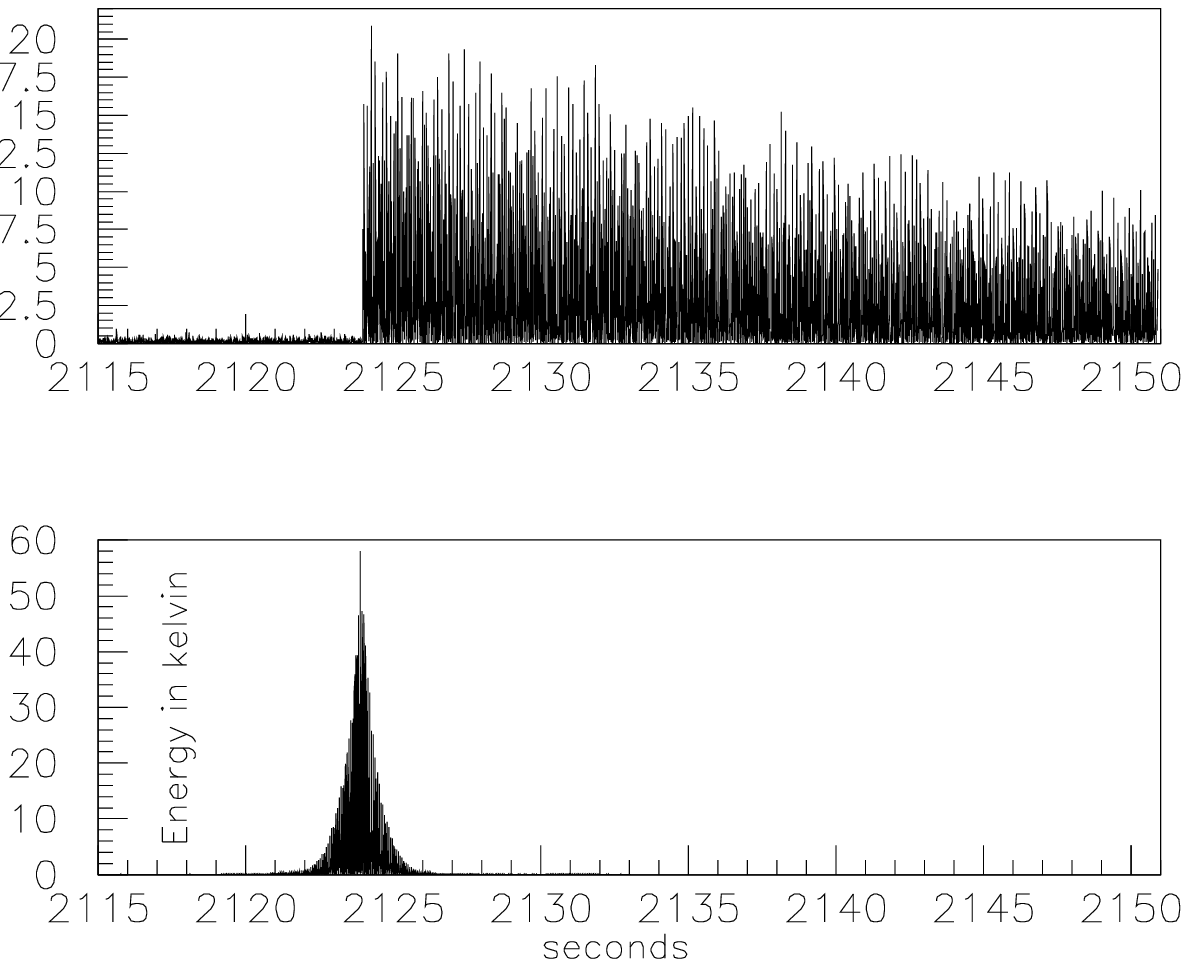}
 \caption{
Time behaviour of the largest coincident NAUTILUS event. In the upper
figure we show the NAUTILUS signal (volt squared) before optimum filtering
versus the UT time expressed in seconds, from the preceding midnight. From the
decay we evaluate the merit factor of the apparatus, Q=1.7 105. The lower plot
shows the data after filtering, in units of kelvin. 
\label{bigone} }
\end{figure}

For a particle density greater than 600 $\frac{particles}{m^2}$
the coincidences reduce to twelve, with n
=0.78. For each coincidence we give the quantity $T_{eff}$, 
the noise of the g.w.
detector during the ten minutes preceding the c.r. event. The time is that
recorded by the c.r. detector.
We notice an unexpected extremely large NAUTILUS event in coincidence with a
c.r. event, with energy  E=57.89 kelvin. Both the up and down particle density
of the c.r. detector are the largest ones in this case. The filtered and
unfiltered data for this event are shown in fig. \ref{bigone}. 
The time of the NAUTILUS event is obtained with good accuracy from the data,
given the very large value of SNR=15860: to=2123.928 s with an error of the
order of 10 ms. The time when the c.r. event has been observed is 2123.9222 s
with a time error of the order of about 1 ms. The difference of 6 ms is within
the experimental error of the g.w. time events (at present our time accuracy for
the NAUTILUS apparatus has been since improved).

\section{Discussion}
\begin{figure}
 \vspace{9.0cm}
\includegraphics{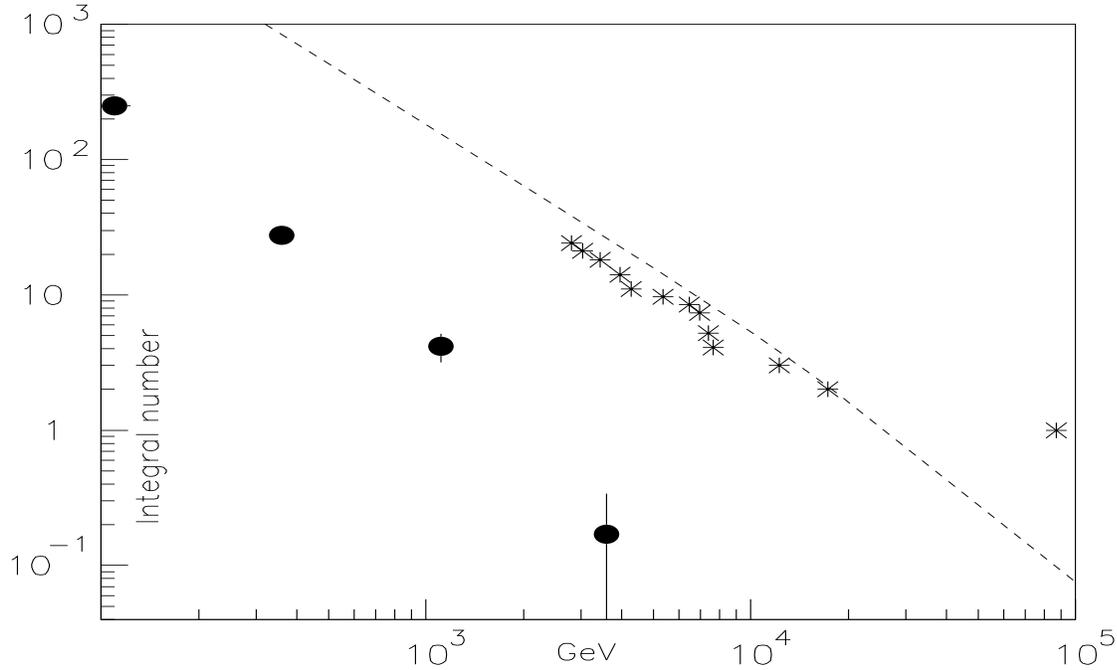}
 \caption{
Comparison between calculations and measurements. The asterisks indicate
the integrated number of coincident events versus the energy delivered by the
c.r. to the bar, expressed in GeV units, to be compared with the points having
error bars, which give the number of events due to hadrons, we expect in the
NAUTILUS bar. The dashed line is the experimental integral spectrum for the
hadronic component of the showers, for the 83.4 days of observation, obtained by
the Cascade experiment. See text.
        \label{adroni} }
\end{figure}

We have found coincidences between NAUTILUS events and c.r. showers. Using eq.
\ref{evw}
we find that the largest NAUTILUS event requires that W=87 TeV of energy be
released by the shower to the bar. There are several points, which must be
clarified and discussed:\\
1.~~~~~Using the down particle
 density shown in Table \ref{eventi} we can calculate the energy
of the NAUTILUS signals that we expect under the hypothesis the shower consists
of electrons. In the previous work \cite{prl},
 finalized to the study of small signals, we
had found that this energy is given by $E=\Lambda^2~4.7~10^{-10}~kelvin$
 where $\Lambda$ is the
number of particles in the bar. For the biggest event the above formula gives
E=0.019 K, that is more than three orders of magnitude smaller than the recorded
58 K. In the same way we calculate energies much smaller than those reported for
all the coincident events of Table 1. Thus we conclude that all, or most of, the
observed NAUTILUS events are not due to electromagnetic showers. On the
contrary, when using the NAUTILUS measurements at zero time delay with energy of
the order or below the noise and add them up at the cosmic ray trigger time, as
done in the previous analysis, we find that the electromagnetic showers account
for the energy observations within a factor of three. For the previous result
\cite{prl}
the energy of the small signals is correlated with the c.r. particle density.
Instead no correlation with the lower particle density is found for the eighteen
large signals given in Table \ref{eventi}.
\begin{figure}
 \vspace{9.0cm}
\includegraphics{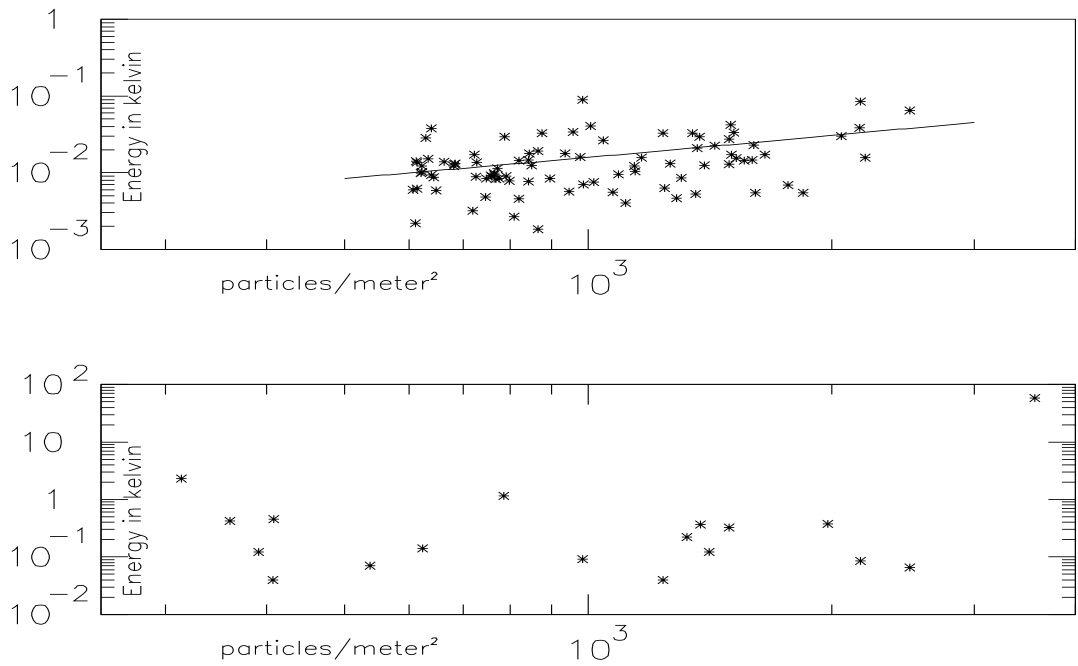}
 \caption{
Correlation between the NAUTILUS signals and the c.r. particle density.
The upper graph shows the correlation of the NAUTILUS energy at zero delay
(respect to the c.r. events) versus the corresponding c.r. lower particle
density, for the 92 data points considered in the previous analysis. The
correlation coefficient is 0.30, with a probability to be accidental of less
than 1\%. If we eliminate the three largest data points with energy greater than
100 mK, which belong also to the family of events of Table 1, the correlation
coefficient increases to 0.42 with 89 data points, with a probability smaller
than $10^{-4}$
 for the correlation to be accidental. Instead the lower plot shows no
correlation between the energy of the NAUTILUS coincident events analysed in
this paper and the corresponding c.r. particle density.
\label{correla} }
\end{figure}
This is shown in fig.\ref{correla},
 and it confirms the idea that the observed large events
are not due to electromagnetic showers. In conclusions, the NAUTILUS signals are
associated to two distinct  families of c.r. showers. In one family the signals
can be interpreted as due to the electromagnetic component of the showers, in
the other family the known c.r. particles in the shower do not justify the
amplitude or the rate of the observed signals.\\
2.~~~~~One must consider the possibility  that the large events are due to the
contribution of hadrons in the showers \cite{sih}.
 Previous calculations have been
made \cite{cocco,peter}
 on the frequency of both hadrons and multihadrons showers. The
calculated values appear to disagree with our observation by more than an order
of magnitude. Recently we have estimated the expected rate of hadronic events in
the bar by means of new Monte Carlo calculations, using the CORSIKA package
\cite{heck}
with the QGSJET model for the hadronic interaction and simulating the NAUTILUS
detector with the GEANT package. This is compared with the integrated number of
coincidences, shown in Table \ref{eventi},
 versus the NAUTILUS event energy. (The covered
time periods are different for the various energy thresholds, which vary during
the observations, depending on the noise. We have normalized the number of
detected events to the total time of 83.4 days). Using eq.\ref{evw}
 we can express the
integral number in terms of the energy $W$ delivered to the NAUTILUS bar by the
cosmic rays. The result is shown in fig. \ref{adroni}. 

In this figure we also report recent
measurements \cite{hora}
 of the hadronic components of extensive air showers, number of
hadronic showers versus their total energy measured with usual particle
detectors. The comparison of these measurements with the result of the Monte
Carlo calculation shown in fig. \ref{adroni} with the error bars prove that the
calculations have been done correctly, since, because of the small diameter of
the bar, we expect that only a few percent of the hadronic energy is absorbed by
the bar, just as shown in fig.\ref{adroni}.

An immediate finding is that the highest energy event occurs in a time period
more than one hundred times shorter than estimated under the hypothesis that the
signals in the bar are due to hadrons. This big specific event could be
explained as due to a large fluctuation, but we also notice a large disagreement
between predicted and observed rates for all other events. Thus our observations
exceed the expectation by one or two orders of magnitude.\\
3.~~~~~We must also consider the possibility that formula 
\ref{evw} does not always apply,
either because the Gr$\ddot{u}$neisen
 coefficient might be larger at the temperature of
NAUTILUS when the Aluminium is superconductor and the specific heat approaches
rapidly zero, or because the impact of a particle could trigger non-elastic
audiofrequency vibrational modes with a much larger energy release. This has
been already suggested \cite{fitz}
  for the case of the interaction with gravitational
waves, to explain cross-sections possibly higher than calculated. However, in
this case, the agreement we have found for the small signals between experiment
and calculation using eq. \ref{evw}
 requires that the breaking of the model occur rather infrequently. \\
4.~~~~~Other possibilities to explain our observations must be considered, as
anomalous composition of cosmic rays (the observed showers might include other
particles, for instance massive nuclei or exotic particles like
nuclearites \cite{deru,witten,nucleariti}
 or Q-balls \cite{cole}). Suggestions for explaining these observations
are solicited.

Finally we remark that the presence of signals due to c.r. does not jeopardize a
coincidence experiment with two or more g.w. detectors. Even without the use of
veto systems employing c.r. detectors, the few dozen of events in a file, which
includes thousand events, does not appreciably affect the number of accidental
coincidences.

\section{Acknowledgements}
Acknowledgment. We thank G. Battistoni for very useful discussions. We also
thank F. Campolungo, R. Lenci, G. Martinelli, E. Serrani , R. Simonetti and F.
Tabacchioni for precious technical assistance.

%
%\section{References}
%

%
\end{document}